\documentclass{aa}
\usepackage{graphicx}

\begin{document}

        \title{The highly ionized disk wind of GRO~J1655-40\thanks{Based on observations obtained with XMM-Newton, 
an ESA science mission with instruments and contributions directly funded by ESA Member States and NASA}}

        \author{G.~Sala\inst{1}, J.~Greiner\inst{1}, J.~Vink\inst{2}, F.~Haberl\inst{1},
		E.~Kendziorra\inst{3} and X.~L.~Zhang\inst{1}
		}
	\institute{Max-Planck-Institut f\"ur extraterrestrische Physik, Postfach 1312, D-85741 Garching, Germany\\
			\email{gsala@mpe.mpg.de}
		\and
		Astronomical Institute, University Utrecht, P.O. Box 80000, NL-3508 TA Utrecht, The Netherlands
		\and
		Institut f\"ur Astronomie und Astrophysik, Abteilung Astronomie, Sand 1, D-72076 T\"ubingen, Germany
		}

        \offprints{G. Sala}

        \date{Received ... /accepted ...}

 
  \abstract
    {The galactic superluminal microquasar GRO~J1655-40 started a new outburst in February 2005, after 
	seven years in quiescence, rising to a high/soft state in March 2005. In this paper we study the 
	X-ray spectra during this rise.
	We observed GRO~J1655-40 with XMM-Newton, on 27 February 2005, in the low/hard state, and 
	on three consecutive days in March 2005, during the rise of the source to its high/soft state.
        The EPIC-pn camera was used in the fast-read Burst mode to avoid photon pile-up.
    	First, we contributed to the improvement of the calibration of the EPIC-pn, 
	since the high flux received from the source required some refinements in the correction of the Charge 
	Transfer Efficiency of the camera.
	Second, we find that the X-ray spectrum of GRO~J1655-40 is dominated in the high/soft state 
	by the thermal emission from the accretion disk, 
	with an inner radius of 13--14$\left[\rm{D}/3.2~\rm{kpc}\right]$~km
	and a maximum temperature of 1.3~keV.  
	Two absorption lines are detected in the EPIC-pn spectra, 
	at 6.7--6.8 and 7.8--8.0~keV, which can be identified either as blended Fe~XXV and Fe~XXVI K$\alpha$ 
	and K$\beta$ lines, or as blueshifted Fe~XXV.
	We find no orbital dependence on the X-ray properties, 
	which provides an upper limit for the inclination of the system of 73$\degr$.
	The RGS spectrometers reveal interstellar absorption features 
	at 17.2~$\rm\AA$, 17.5~$\rm\AA$ (Fe L edges) and 23.54~$\rm\AA$ (OI K$\alpha$).
	Finally, while checking the interstellar origin of the OI line,	we find
	a general correlation of the OI K$\alpha$ line equivalent width with the hydrogen column density using 
	several sources available in the literature. 
       
	}

        \titlerunning{The disk wind of GRO~J1655-40}
        \authorrunning{G. Sala~et~al.}
        \maketitle

%

\section{Introduction}

Galactic microquasars are accreting binary systems ejecting jets at relativistic velocities. 
Both black holes as well as neutron stars have been identified as
the compact, accreting object. The analogy of these systems 
to extragalactic quasars and Active
Galactic Nuclei (AGN) make them excellent laboratories for the
study of the physics involved in accretion disks and ejection
of relativistic jets associated with accreting black holes 
(Mirabel~et~al.~\cite{mir92}). 
The physics ruling black hole systems is essentially the same
in galactic microquasars, hosting stellar black holes, and in
AGN, with supermassive black holes, but the differences in time
scales makes the study of some aspects in microquasars easier
than in AGN. The characteristic timescale for the flow of matter
onto a black hole is proportional to its mass, being of order of minutes
in a microquasar of a few solar masses, but of thousands of
years in a massive black hole of $10^{9}\rm M_{\odot}$. 
In addition,  thanks to their proximity, microquasar jets have proper motions in the plane of sky with
velocities about a thousand times faster than AGN, and two-sided jets can be observed.

The microquasar GRO\,J1655-40 (X-ray nova Sco\,1994, Zhang~et~al.~\cite{zha94}) 
was the second superluminal source discovered in our Galaxy, 
after GRS\,1915+105 (Mirabel~\&~Rodriguez~\cite{mir94}).
The two sources may also be peculiar in the sense that they show evidence
suggesting that both systems contain a maximally spinning black hole (Zhang~et~al.~\cite{zha97}).
Radio images of GRO\,J1655-40 showed twin jets with apparent superluminal motion 
(Tingay~et~al.~\cite{tin95}) moving in opposite directions at 0.92c, and 
the distance was determined to be 3.2$\pm$0.2~kpc (Hjellming~\&~Rupen~\cite{hje95}). 
Using the dust scattering halo observed by ROSAT, Greiner~et~al.~(\cite{gre95}) determined 
a distance of 3~kpc, compatible with the previous determination.
Optical observations in 1996 provided an inclination angle of 69.5$\pm$0.08$\degr$,
a radial velocity semiamplitude of 228.2$\pm$2.2~km~s$^{-1}$ and 
a dynamical mass of the primary component of 7.02$\pm$0.22~M$_{\odot}$ 
(Orosz~\&~Bailyn~\cite{oro97}), indicating that it is a black hole.
Hubble Space Telescope observations showed that the system moves 
in an eccentric orbit with a runaway space velocity (i.e., with respect to the Galactic 
rotation corresponding to its position in the Galactic plane) of 112$\pm$18~km~s$^{-1}$ 
(Mirabel~et~al.~\cite{mir02}), which together with abundance anomalies found in the 
atmosphere of the donor star (Israelian~et~al.~\cite{isr99}) provide evidence for
a supernova origin of the black hole in GRO~J1655-40.

ASCA observations of GRO~J1655-40 in August 1994 and August 1995 provided the first 
detection of absorption lines in an accretion powered source (Ueda~et~al.~\cite{ued98}).
The energy of the lines was found to depend on the X-ray intensity, 
being 6.95~keV (Fe~XXVI K$\alpha$) at 2.2 Crab, and 6.63 and 7.66~keV
(Fe~XXV K$\alpha$ and K$\beta$) at 0.27--0.57 Crab,   
revealing the presence of a highly ionized absorber.
Similar absorption features were also detected for GRS 1915+105 (Kotani~et~al.~\cite{kot00}).
During the outburst in 1997, ASCA observed GRO~J1655-40 
between 25 and 28 February, when it was at 1.1~Crab, and 
revealed an absorption feature at 6.8~keV, interpreted as the blend of the two resonance 
K$\alpha$ lines of Fe~XXVI and Fe~XXV (Yamaoka~et~al.~\cite{yam01}).
Simultaneously, RXTE observations performed on 26 February 1997 
showed emission lines at 5.85 and 7.32~keV, 
interpreted as the first red- and blueshifted Fe~K$\alpha$ 
disk line in a Galactic source (Balucinska-Church~\&~Church~\cite{bal00}), 
indicating the presence of significantly ionized material in a region of the disk 
at $\sim10$ Schwarzschild radii.

\begin{figure}
\resizebox{\hsize}{!}{\includegraphics{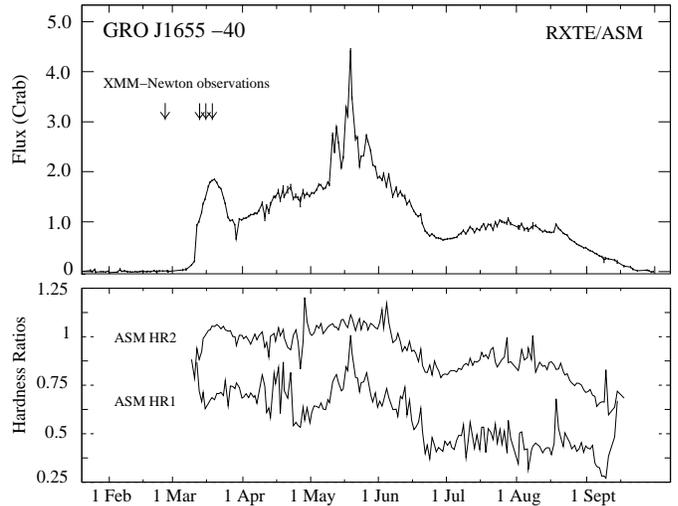}}
\caption{RXTE/ASM light curve of GRO~J1655-40, with dates
of our XMM-Newton observations, and ASM hardness ratios:
HR1, ratio of the ASM count rates (3--5~keV)/(1.5--3~keV), and HR2, (5--12~keV)/(3--5~keV).
Data from the quick-look results provided by the ASM/RXTE team.
\label{fig_lc}}
\end{figure}

The first observed X-ray outbursts of GRO~J1655-40 occurred in 1994--1995  
(Zhang~et~al.~\cite{zha94}, Harmon~et~al.~\cite{har95}), and a 16-month-long outburst 
began on 25 April 1996 (Sobczak~et~al.~\cite{sob99}).
After 7 years of inactivity, 
GRO~J1655-40 left quiescence again on 17 February 2005 (Markwardt~\&~Swank~\cite{mar05}).
The X-ray evolution was followed with RXTE/ASM (Homan~et~al.~\cite{hometal05}) 
and Swift (Brocksopp~et~al.~\cite{bro06}).
GRO~J1655-40 entered in February a low/hard state (Homan~\cite{hom05}) until it
experienced a first outburst in March, moving into a high/soft state and 
reaching $\sim2$ Crab (Fig.~\ref{fig_lc}). The decay of the first 
outburst was followed by a month and a half of increasing 
X-ray flux and finally a strong outburst in very high state in 
May 2005, when the source reached more than 4 Crab. 

Here we present XMM-Newton observations performed in February-March 2005, 
aimed at obtaining detailed spectroscopy during the rise to the high/soft state.
We use the canonical distance of 3.2~kpc throughout (see note added in proof, however).


\section{Observations and data reduction}

We observed GRO~J1655-40 with XMM-Newton some days before the rise to the soft/high state, on
27 February 2005 (40 ks; obs. id. 0112921-301), and again on 14, 15 and 16 March 2005,
close to the maximum of the first outburst (15 ks each, obs. id. 0112921-401/-502/-601; 
see Fig.~\ref{fig_lc} and Table~\ref{tab_obs}).
The EPIC-pn camera was used in the Burst mode, a fast read-out mode
that permits the observation of very bright sources with a high time resolution (Kuster~et~al.~\cite{kus99}). 
In the Burst mode, only one of the 12 CCDs of the camera is used. The electrons released
in the area of the CCD where the source is positioned are quickly shifted towards the read-out node to
prevent pile-up, and then normally read out. In this mode, the spatial information 
in one dimension (shift direction) is lost 
and only 3\% of the total time is effectively used, but bright sources can be observed 
without photon pile-up problems. 
The two MOS cameras were switched off for telemetry limitations, while the RGS
spectrometers were operating in normal spectroscopy mode.

XMM-Newton data were reduced using SAS 6.1, and XSPEC 11.3 was used for spectral analysis.
Solar flares affected partially our observations, and some time intervals could not be used for analysis
(in particular, the second half of the first observation lasting 20 ks, 
the first 2 ks of the second and the first 5 ks 
of our last observation). For the EPIC-pn data, only single events
with energy higher than 0.4~keV were taken into account. Events with RAWY$\geq$140 were
excluded to avoid direct illumination by the source, and the source spectra were
extracted from a box including 10 RAWX columns at each side of the source position and 
ignoring also RAWY$\leq$10. 
The high luminosity of GRO~J1655-40 at the time of our XMM-Newton observations (almost twice
as bright as the Crab, the source used for Burst mode calibration)
made evident a rate-dependence of the Charge Transfer Efficiency (CTE) 
not included in the calibration. We determined
this dependence and corrected our spectra accordingly (see appendix).
 
In the RGS spectra, the energy range 0.3--2.0~keV was initially included.
We applied a preliminary version of wavelength and time dependent corrections to the RGS 
effective area, which will be implemented in the next release of the XMM-Newton software.
Nevertheless, during our three March observations, strong differences were found between the responses of the 
different nodes of the RGS, probably due to the high flux of the source. 
We have thus ignored the energy ranges with evident problems, more precisely,
energies above 1~keV for the first order in all cases, and between 0.85 and 1.2~keV in the second order of the RGS2
on the 14th March observation. Since both first and second order spectra 
of the two RGS are available, no energy range remains uncovered despite the rejection of these data.

Simultaneous RXTE observations of GRO~J1655-40 are publicly available (Table~\ref{tab_obs}). 
We have included the standard spectral products from the PCA and HEXTE instruments 
in our analysis, to constrain the model 
components at hard X-rays. When EPIC-pn and PCA spectra of March observations 
are attempted to be fit simultaneously, large 
residuals appear for the PCA spectrum for energies lower than $\sim6$~keV, in the region of the 
instrumental Xenon edges. Since the accuracy of the PCA response function is limited by the poor 
modeling of these edges (Jahoda~et~al.~\cite{jah05}), it is possible that the high flux of the 
source in these observations is revealing some inaccuracy in the calibration, as in the 
case of the XMM-Newton instruments. We thus only consider the PCA spectrum for 
energies above 6~keV.

Neither orbital variations, like eclipses or periodic modulations, nor dips 
are observed in the light curve, and the flux of the 
source remains constant within 30\%.


\begin{table*}
\centering
\caption{\label{tab_obs} Observation log}
\begin{tabular}{ c c c c c c c }
\hline\hline
\noalign{\smallskip}
Instrument & Date & Start & End & Exposure & Energy range & Flux ($\times 10^{-10}$) \\
& & (UTC) & (UTC) & (ks) & (keV) & (erg s$^{-1}$cm$^{-2}$)\\ 
\noalign{\smallskip}
\hline 
\noalign{\smallskip}
XMM/EPIC-pn &   27-02-2005 & 07:56 & 19:34 & 41.8  & 0.4--10.0 & 2.5\\
XMM/RGS1 \& 2, order 1 &27-02-2005 & 07:46 & 19:35 & 42.5  &  0.3--2.0 & 0.5\\
XMM/RGS1 \& 2, order 2 &27-02-2005 & 07:46 & 19:35 & 42.5 & 0.54--2.0 & 0.4\\
RXTE/PCA &         27/28-02-2005   & 20:37 & 02:53 & 13.0  & 6.0--25.0 & 5.9\\
RXTE/HEXTE &  27/28-02-2005 	   & 20:37 & 02:17 & 4.2   &  20--200  & 21\\
\noalign{\smallskip}
XMM/EPIC-pn & 14-03-2005 & 16:27 & 20:36 & 14.9  & 0.4--10.0 & 214\\
XMM/RGS1 \& 2, order 1 & 14-03-2005 & 16:17 & 20:37 & 15.6  &  0.3--1.0 & 2.9\\
XMM/RGS1 \& 2, order 2 & 14-03-2005 & 16:17 & 20:37 & 15.6  & 0.48--1.8 & 40\\
RXTE/PCA    & 14-03-2005 & 13:02 & 19:17 & 12.9  & 6.0--25.0 & 54\\
RXTE/HEXTE  & 14-03-2005 & 13:02 & 19:17 &  4.0  &  20--200  & 36\\
\noalign{\smallskip}
XMM/EPIC-pn & 15-03-2005 	   & 16:27 & 20:36 & 14.9  &  0.4--10.0 & 208\\
XMM/RGS1 \& 2, order 1 &15-03-2005 & 16:17 & 20:37 & 15.6  &  0.3--1.0 & 2.9\\
XMM/RGS1 \& 2, order 2 & 15-03-2005 & 16:17 & 20:37 & 15.6   & 0.48--1.8 & 40\\
RXTE/PCA    & 15-03-2005 & 14:08 & 19:26 & 11.6  & 6.0--25.0 & 48 \\
RXTE/HEXTE  & 15-03-2005 & 14:08 & 19:36 &  3.5  &   20--200 & 16 \\
\noalign{\smallskip}
XMM/EPIC-pn & 16-03-2005 & 16:14 & 20:23 & 14.9  & 0.4--10.0 & 260\\
XMM/RGS1 \& 2, order 1 & 16-03-2005 & 16:04 & 20:24 & 15.6  & 0.3--1.0  & 3.2 \\
XMM/RGS1 \& 2, order 2 & 16-03-2005 & 16:04 & 20:24 & 15.6  & 0.54--1.8 & 45 \\
RXTE/PCA    & 16-03-2005 & 17:08 & 22:37 & 11.1  & 6.0--25.0 & 70 \\
RXTE/HEXTE  & 16-03-2005 & 17:08 & 22:37 &  3.5  &  20--200  & 34 \\
\noalign{\smallskip}
\hline 
\noalign{\smallskip}
\end{tabular}
\end{table*}

\section{Spectral analysis}

\subsection{February observation}

During the first observation, on 27 February 2005, before the start of the first outburst, 
the source was in a low/hard state, with the XMM-Newton plus RXTE instruments showing a 
spectrum dominated by an absorbed power law with photon index $\Gamma$=1.48$\pm$0.01 
(uncorrelated 90\% error for 1 degree of freedom).
This is in agreement with the results from RXTE observations starting only two days later,
in the period 1--6 March 2005 ($\Gamma$=1.4--1.6, Swank~\&~Markwardt~\cite{swa05}),
and about 10\% flatter than reported by Brocksopp~et~al.~(\cite{bro06}) from Swift observations on 6 March 2005
($\Gamma$=1.72$\pm$0.03). 
The fit with a power law leaves excess residuals at low energies both in the 
EPIC-pn and the RGS spectra, but the fit improves when the spectrum is modified 
by the scattering of dust in the line-of-sight. Greiner~et~al.~(\cite{gre95})
found in ROSAT HRI observations in 1994 a halo around GRO~J1655-40 due 
to the scattering of the X-rays by interstellar dust. 
We modify thus our spectral model with the 'dust' model available in XSPEC,
which assumes that the scattered flux appears as a uniform disk on the sky
(as for a source constant in time) whose size
has a 1/E dependence and whose total flux has a 1/E$^2$ dependence.

The RXTE/PCA spectrum also shows residuals between 6--7~keV, which are well 
fitted with an iron fluorescence line with energy fixed at 6.4~keV, 
for which we derive an equivalent width of 135$\pm$20~eV. 
The line is not seen in the EPIC-pn data, and fixing the energy at 
6.4~keV, we obtain an upper limit of 140~eV for the equivalent width.
The presence of an iron fluorescence line in the low/hard state is confirmed 
by Swift observations one week after our XMM-Newton observation
(Brocksopp~et~al.~\cite{bro06}).
All parameters of the fit for this first observation are listed in
Tables~\ref{tab_mod} and \ref{tab_xte}.

\begin{figure}
\resizebox{\hsize}{!}{\includegraphics{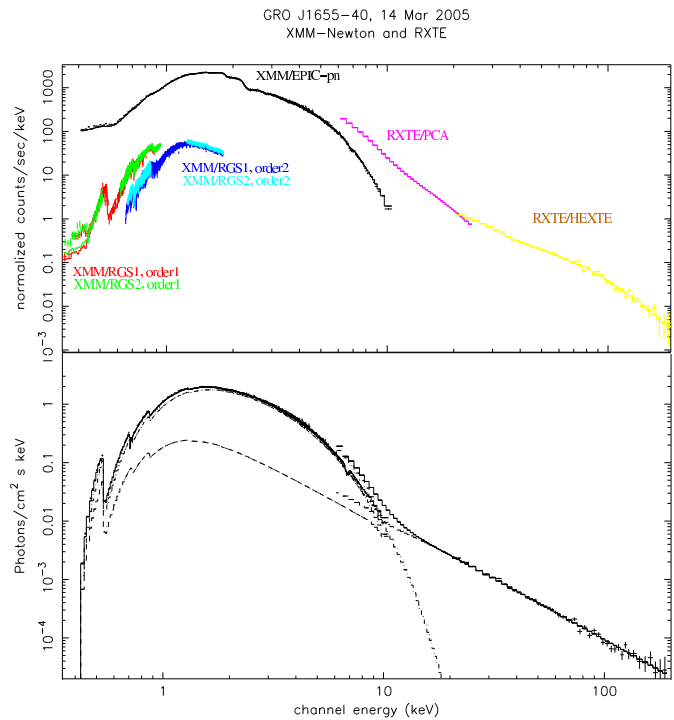}}
\caption{XMM-Newton and RXTE observed spectra (upper panel), and unfolded spectra showing the contribution 
of the thermal disk emission (soft energies) and the power law (hard energies) to the total spectrum (lower panel).
Residuals for the three March observations are shown in Figs.~\ref{fig_oxygen}, 
\ref{fig_rgs} and \ref{fig_pn}.}
\label{fig_all}
\end{figure}

\subsection{March observations}

The three March observations have very similar spectra, with the XMM-Newton 
continuum spectra dominated by an absorbed multicolor disk (MCD), 
with kT$_{\rm{in}}$ increasing with time from 1.25 to 1.35~keV, 
and the higher energy spectra from RXTE represented by a soft power law, with $\Gamma$=2.1--2.3~keV
(see Fig.~\ref{fig_all} and Table~\ref{tab_mod}). 
The simultaneous fit of the EPIC-pn, RGS, PCA and HEXTE data 
with the MCD plus power law model leaves large residuals again below 0.6~keV in both the EPIC-pn and 
the RGS spectra, as well as in the range 8--15~keV in the XTE/PCA spectra.

The soft excess in the XMM-Newton instruments, 
as in the case of the February observation, is most probably related to the dust scattering. 
Although the global fit improves after adding the dust halo effect, the RGS spectra still 
show an excess for energies lower than 0.5~keV. 
Using different abundance tables or leaving the abundances as free 
parameters did not help to fit this soft excess. 
This is probably related to the 
more limited sky area included in the RGS than in the EPIC-pn extraction regions. 
A detailed analysis of the halo, including spatially resolved
spectroscopy, is currently underway and will be reported separately.
As a first approximation, fixing the dust halo size to 10 times the beam size,
from the fit with the EPIC-pn data we obtain in all cases 
a scattering fraction at 1~keV of 17--20\%, similar to the 21\% relative intensity of the halo found by 
Greiner~et~al.~(\cite{gre95}).

The quality of the fit in the 8--15~keV energy range spectra of the RXTE instruments
during the three March observations improves with the addition of a smeared edge at 8.6--8.8~keV
(see Table~\ref{tab_xte}).
An iron edge at 8.81~keV was also detected in the 1994 outburst from ASCA observations (Ueda~et~al.~\cite{ued98})
and at 8~keV with RXTE observations during the 1996--1997 outburst (Sobczak~et~al.~\cite{sob99}).
Nevertheless, for the 15 and 16 March observations, even after including the edge, the fit 
of the PCA spectra below 20~keV is poor. It is worth noticing that Sobczak~et~al.~(\cite{sob99})
did also not find a good fit for the PCA spectra between 2.5 and 20~keV
with MCD plus power law model, neither including 
emission lines, edges nor other effects, for the observations performed 
during the rise of the source to the first burst in 1996, in 
a phase of the light curve similar to our March observations.

\begin{table*}
\centering
\caption{\label{tab_mod} Continuum Spectral Model for XMM-Newton and RXTE observations}
\begin{tabular}{ c c c c c }
\hline \hline
\noalign{\smallskip}
Parameters & 27 Feb 2005 & 14 March 2005 & 15 March 2005 & 16 March 2005 \\
\noalign{\smallskip}
\hline
\noalign{\smallskip}
N$_{\rm {H}} (\times 10^{21})$ cm$^{-2}$& 5.2$\pm$0.2 	& 5.46$\pm$0.02	   & 5.57$\pm$0.02 	& 5.45$\pm$0.03\\
Dust fraction$^{\mathrm{a}}$ 		& 0.15$\pm$0.03 & 0.186$\pm$0.002  & 0.169$\pm$0.003 	& 0.184$\pm$0.003\\
kT$_{in}$ (keV)				&     -		& 1.231$\pm$0.003  & 1.268$\pm$0.003 	& 1.326$\pm$0.003 \\
R$_{in}^{\mathrm{b}}$ (km) 				&     -		& 14.25$\pm$0.05   & 13.25$\pm$0.05 	& 13.55$\pm$0.05 \\
$\Gamma$ 				& 1.48$\pm$0.01 & 2.10$\pm$0.01    & 2.27$\pm$0.01 	& 2.13$\pm$0.01\\
K$^{\mathrm{c}}$ (ph~keV$^{-1}$cm$^{-2}$s$^{-1}$@ 1~keV)& 0.0385$\pm$0.0007 & 0.94$\pm$0.02 & 0.76$\pm$0.07   	& 1.05$\pm$0.02 \\
L$_{~0.4--10~\rm{keV}} ^{\mathrm{d}}$ (erg\,s$^{-1}$)&3.7$\times10^{35}$&3.4$\times10^{37}$&3.4$\times10^{37}$&4.2$\times10^{37}$ \\
\noalign{\smallskip}
\hline
\end{tabular}
\begin{list}{}{}
\item[$^{\mathrm{a}}$] Dust scattering fraction at 1~keV. 
Halo size is fixed to 10 times the detector beam-size at 1~keV.
\item[$^{\mathrm{b}}$] For a distance of 3.2 kpc and an inclination of 70$\degr$.
\item[$^{\mathrm{c}}$] Normalization constant for the power law component.
\item[$^{\mathrm{d}}$] Luminosity in the range 0.4--10~keV, for a distance of 3.2 kpc.
\end{list}
\end{table*}

\begin{table*}
\centering
\caption{\label{tab_xte} Features in RXTE/PCA spectra}
\begin{tabular}{c c c c c}
\hline\hline
\noalign{\smallskip}
Parameters & 27 February 2005 & 14 March 2005 & 15 March 2005 & 16 March 2005 \\
\noalign{\smallskip}
\hline
\noalign{\smallskip}
Line energy (keV) (FIXED) 		& 6.4 		& - & - & - \\
1$\sigma$ line width (eV)		& 300$\pm$100 	& - & - & - \\
Equivalent width (eV)			& 135$\pm$20	& - & - & - \\
Identification 				& Fe I 		& - & - & - \\
\noalign{\smallskip}\noalign{\smallskip}
Edge energy (keV)  & - 	& 8.68$\pm$0.05	& 8.57$\pm$0.03	& 8.77$\pm$0.04\\
Edge optical depth & - 	& 0.44$\pm$0.05	&  0.5$\pm$0.2	& 0.57$\pm$0.05\\
Edge width 	   & - 	&  1.2$\pm$0.2 	&  1.9$\pm$0.5 	& 1.3$\pm$0.2\\
\noalign{\smallskip}
\hline
\end{tabular}
\end{table*}

\subsection{Absorption features in the March observations}

\begin{figure}
\resizebox{\hsize}{!}{\includegraphics{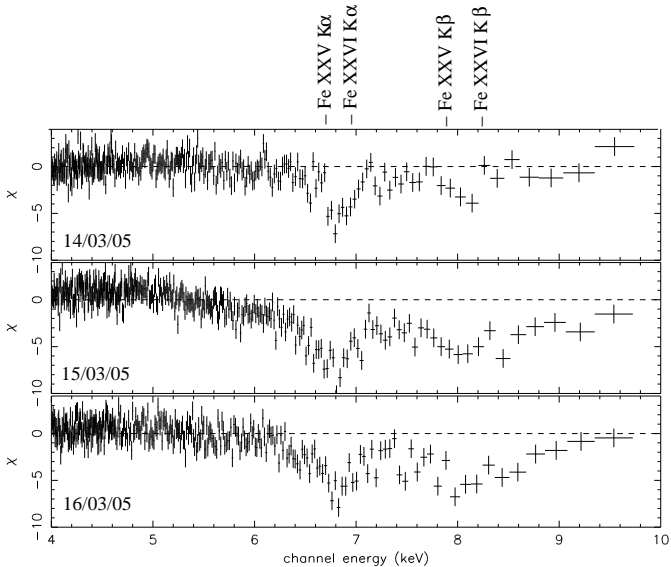}}
\caption{EPIC pn residuals of GRO~J1655-40 spectra of our three March 2005 observations 
at zero velocity, after fitting an absorbed multi-temperature disk model.}
\label{fig_pn}
\end{figure}

Clear structured residuals appear also in the EPIC-pn spectrum, at $\sim6.8$ and  $\sim7.9$~keV, 
corresponding to absorption features in the Fe K region (see Fig.~\ref{fig_pn} and Table~\ref{tab_epic}).
In all three March observations the energy of the lines is the same, 
corresponding either to blended Fe~XXV and Fe~XXVI K$\alpha$ and K$\beta$ lines, or
to blueshifted Fe~XXV lines. In the latter case, given the energy resolution of the 
EPIC-pn (150~eV @ 6.4~keV), the velocity is not well constrained.

\begin{figure}
\resizebox{\hsize}{!}{\includegraphics[angle=-90]{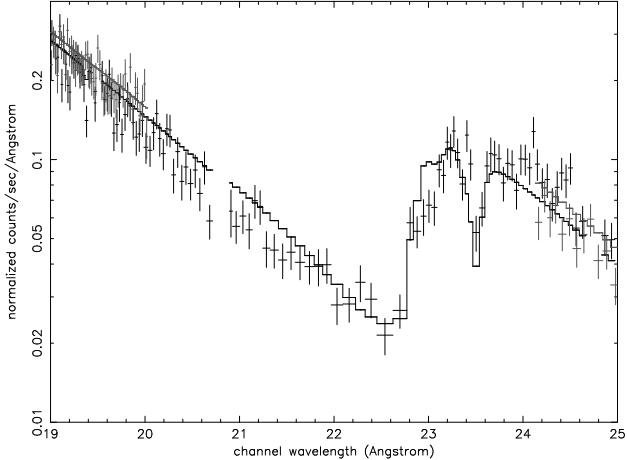}}
\caption{RGS1 (black) and RGS2 (grey, contributing only below 20$\rm \AA$ and above 24$\rm \AA$.) 
around the oxygen edge in the first order spectra of the 14th March 2005 observation, 
fitted with the new TBabs model.}
\label{fig_oxygen}
\end{figure}

\begin{figure}
\resizebox{\hsize}{!}{\includegraphics[angle=-90]{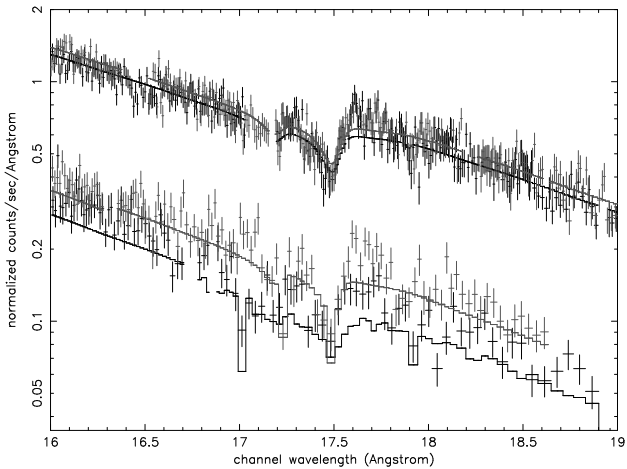}}
\caption{First and second order (with lower count rate) spectra of GRO~J1655-40 in 
the range 16--19$\rm \AA$, obtained with RGS1 (black) and RGS2 (grey) on 14 March 2005, showing 
the ISM Fe L edges fitted with the new TBabs model.}
\label{fig_rgs}
\end{figure}

In the three March observations, the fit of the RGS spectra 
around the interstellar neutral oxygen edge at 22.6--23.2 $\rm \AA$ (534--549~eV) with 
the \emph{phabs} absorption model is 
unacceptable unless the ``Vern'' cross-sections (Verner~et~al.~\cite{ver96}) available
in XSPEC are used. Even in this case, the fit leaves clear residuals,
indicating the presence of an additional absorption line at 23.5 $\rm \AA$ (Fig.~\ref{fig_oxygen}),
which corresponds to the 1s-2p interstellar absorption line of neutral oxygen.
Given the energy resolution of the RGS, the observed maximum width of the line is compatible with it being 
a narrow line originating in the interstellar medium. 
The RGS spectra show also clear absorption features at 17.2~$\rm \AA$ and 17.5~$\rm \AA$  (Fig.~\ref{fig_rgs}),
which correspond to Fe L$_{2,3}$ (2p$_{1/2}$, 2p$_{3/2}$) edges, 
also observed in the interstellar absorption towards some other 
sources (Wilms~et~al.~\cite{wil00}, Paerels~et~al.~\cite{pae01}, Schulz~et~al.~\cite{sch02}). 
As in these previous cases, simple edge models do not provide a good fit for these features.

Since both the line at 23.5 $\rm \AA$ and the edges at 17.2~$\rm \AA$ and 17.5~$\rm \AA$ are caused 
by interstellar gas but are not well reproduced by the simple \emph{phabs} model in XSPEC,
the recent new version of the \emph{TBabs} model (Wilms~et~al.~\cite{wil06}, 
http://astro.uni-tuebingen.de/$\sim$wilms/research/tbabs/) has been used.
This has been shown to provide a good fit to the 
complex oxygen edge (Juett~et~al.~\cite{jue04}), including two edges and five absorption lines, 
and to the iron absorption features (Juett~et~al.~\cite{jue06}).
It provides a good fit also for our RGS features (see Figs.~\ref{fig_oxygen} and \ref{fig_rgs}), 
with a hydrogen column N$_{\rm{H}}:7.3(\pm0.3)$, $7.0(\pm0.3)$, $6.9(\pm0.5)\times10^{21}\rm{cm}^{-2}$, 
and an overabundance of oxygen with respect to solar of $1.45(\pm0.15),1.6(\pm0.1),1.45(\pm0.15)$ 
(for days 14, 15 and 16 March 2005 respectively).

\begin{table*}
\centering
\caption{\label{tab_epic} Absorption features in EPIC-pn spectra}
\begin{tabular}{c c c c}
\hline\hline
\noalign{\smallskip}
Parameters & 14 March 2005 & 15 March 2005 & 16 March 2005 \\
\noalign{\smallskip}
\hline
\noalign{\smallskip}
Line energy (keV) 		& 6.82$\pm$0.06		& 6.71$\pm$0.06	& 6.75$\pm$0.05\\
1$\sigma$ line width (eV)	& 110$^{+100}_{-50}$ 	& 200$\pm$100 	& 250$\pm$50\\
Equivalent width (eV) 		& 60$^{+30}_{-10}$ 	& 100$\pm$60	&  110$\pm$30 \\
Identification$^{\mathrm{a}}$	& \multicolumn{3}{c}{Fe~XXV K$\alpha$ (E$_0$ 6.700~keV)/Fe~XXVI K$\alpha$ (E$_0$ 6.952/6.973~keV)} \\
v (km~s$^{-1}$)$^{\mathrm{b}}$		& -5200$\pm$2600 	& -400$\pm$2700	&-2200$\pm$2200\\
\noalign{\smallskip}
Line energy (keV) 		& 8.04$\pm$0.06		& 7.85$\pm$0.15		& 7.9$\pm$0.1\\
1$\sigma$ line width (eV)  	& 100$^{+100}_{-50}$	& 700$\pm$200		& 600$\pm$200\\
Equivalent width (eV)      	&54$^{+1}_{-25}$ 	&  230$^{+20}_{-50}$	&  240$\pm$40\\
Identification$^{\mathrm{a}}$ & \multicolumn{3}{c}{Fe~XXV K$\beta$ (E$_0$ 7.872/7.881~keV)/Fe~XXVI K$\beta$ (E$_0$ 8.246/8.252~keV)}\\
v (km~s$^{-1}$)$^{\mathrm{b}}$	 	& -5900$\pm$2200 	& 1200$\pm$5700	& -700$\pm$3800\\
\noalign{\smallskip}
\hline
\end{tabular}
\begin{list}{}{}
\item[$^{\mathrm{a}}$] Based on CHIANTI data base (Dere~et~al.~\cite{der97}, Landi~et~al.~\cite{lan06})
\item[$^{\mathrm{b}}$] Blueshift with respect to Fe~XXV lines. 
\end{list}
\end{table*}


\section{Discussion}

\subsection{The dust halo}

The effect of the scattering dust halo is clearly detected in the soft X-ray spectra. 
The scattering fraction at 1~keV is changing during the three March observations, 
decreasing from 18.6($\pm$0.2)\% to 16.9($\pm$0.3)\% in the first 24 hours, and then increasing 
again to 18.4($\pm$0.3)\% during the next day. 
This temporal variation is consistent with a constant dust halo but a
variable X-ray source flux: the time needed for the soft photons originating on the disk to 
travel to the scattering site far away from the source causes a delay in the flux variations
of the dust halo with respect to the variations of the source. As a result, just after an increase of the 
disk temperature and thus of the source soft spectrum, the fraction of scattered photons with 
respect to the source is temporally smaller (during the second March observation) until the increasing 
source photons reach the outer part of the halo and their scattered photons also reach the observer, 
increasing again the scattering fraction (third March observation). 
This effect has been known for some time. It was also observed by ROSAT 
and used to estimate the distance to GRO~J1655-40 (Greiner~et~al.~\cite{gre95}).

\subsection{Continuum}

The maximum disk temperature observed in the March observations,  
increasing from 1.25 to 1.35~keV, is consistent with the values commonly 
observed in black hole binaries in high/soft state 
(see for instance McClintock~\&~Remillard~\cite{mcc06}), 
and with the Swift observations of the same 
period (Brocksopp~et~al.~\cite{bro06}). The inner radius,  
\mbox{$\rm R_{in}\sim13-14\left[\rm{D}/3.2\rm{kpc}\right]$~km}, 
is only a little larger than the gravitational radius of the black hole, 
R$_{\rm g}=\rm G\rm M/ \rm c^2=10$~km, consistent with a rapidly spinning black hole 
(Zhang et al. \cite{zha97}).

From the fit parameters of the MCD model
the luminosity of the disk can be calculated as
$\rm L_x=4\pi \rm R_{in}^2\sigma \rm T_{in}^ 4$ (Makishima~et~al.~\cite{mak86}).
With our values, the accretion disk luminosity is 
\mbox{$6-7\times10^{37}\left[\rm{D}/3.2\rm{kpc}\right]^2$~erg~$\rm s^{-1}$}. 

With the accretion disk luminosity, the accretion rate can be 
calculated as $2\rm L_x \rm R_{in}/\rm G \rm M \rm g^2$  (Makishima~et~al.~\cite{mak86}), 
where $\rm g=(1-\rm R_{g}/\rm R_{in})^{1/2}$ is a correction for the general relativity,
R$_{\rm g}$ is the gravitational radius and M is the mass of the black hole.
From the accretion disk luminosity, the disk inner radius and the black hole mass of GRO~J1655-40,
we obtain an accretion rate during the high/soft state of $\sim10^{-8}~\rm M_{\odot}~\rm{yr}^{-1}$.

The rise of the disk blackbody component during the high/soft state in March 
is reflected in the power law index of the corona, which increases from 1.48$\pm$0.01 in February 27 to 
values higher than 2 in the March observations. With the rise of the thermal disk emission, 
the soft luminosity increases and the electrons in the corona are cooled more efficiently, 
resulting in a softer spectrum of the comptonized photons, i.e., a steeper power law.

\subsection{Interstellar absorption}

The interstellar OI~K$\alpha$ line is clearly detected at 23.5~$\rm \AA$ in our three March 2005 RGS spectra. 
This line was also found in the RGS spectra of other sources, and distinguished from the 
instrumental components around the interstellar oxygen edge, at 23.05 and 23.35 $\rm \AA$
(de Vries~et~al.~\cite{cor03}). Juett~et~al.~(\cite{jue04}) detected the same line 
in the high resolution spectra of seven X-ray binaries using the Chandra/HETGS, 
as part of their study of the structure of the oxygen absorption edge caused by the interstellar medium.
More recently, it has also been detected in the Cyg~X-2 spectrum (Costantini~et~al.~\cite{cos05}), 
in the observations that have allowed the first spatially resolved spectroscopic study of a scattering dust halo, 
as well as in the absorption towards LMC~X-3 (Wang~et~al.~\cite{wan05}) 
and the recently discovered black hole candidate XTE~J1817-330 (Sala~\&~Greiner~\cite{sal06}).

\begin{figure}
\resizebox{\hsize}{!}{\includegraphics[angle=0]{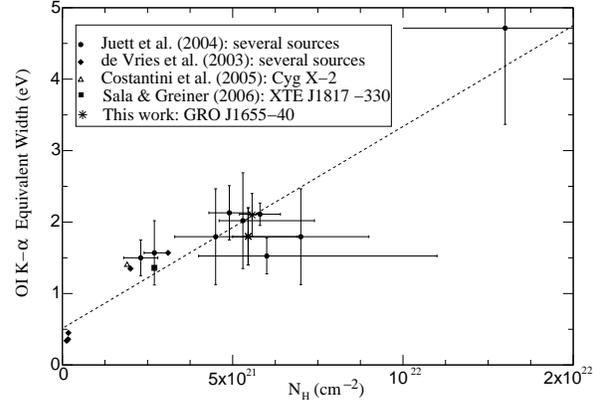}}
\caption{OI K-$\alpha$ line equivalent width versus hydrogen column density
for several galactic sources. The dashed line indicates an approximate linear correlation,
EW(eV)=$0.5+2.8\times10^{-22}\rm N_{\rm H}(\rm{cm}^{-2})$.}
\label{fig_oh}
\end{figure}

To check the interstellar origin of the OI~K$\alpha$ line observed in the RGS spectra 
of GRO~J1655-40, we have looked for a correlation of the equivalent width of the line and 
the hydrogen column density, using the above mentioned works (Fig.~\ref{fig_oh}). 
We find that the equivalent width of the OI~K$\alpha$ line, EW, is indeed correlated
with the column density, N$_{\rm H}$, as EW(eV)=$0.5(\pm0.1)+2.8(\pm0.3)\times10^{22}\rm N_{\rm H}(\rm{cm}^{-2})$.
The values observed for GRO~J1655-40 fall well within this correlation for the ISM. 
We find that an alternative correlation with zero regression constant could be  
EW(eV)=$3.6(\pm1.6)\times10^{22}\rm N_{\rm H}(\rm{cm}^{-2})$, but
this would underestimate the equivalent width for the sources
with hydrogen column density less than $5\times10^{21}\rm{cm}^{-2}$.

\subsection{Orbital dependence and inclination of the binary system}

Since the orbital period of GRO~J1655-40 is $\sim2.6$ days and our three March observations
were taken with one day intervals, we are covering approximately one whole orbital cycle. 
We can extrapolate the Orosz~\&~Bailyn~(\cite{oro97}) ephemeris, to obtain the 
orbital phase of our observations (error is 0.08 in all cases):
0.31--0.40 (27 February), 0.17--0.24 (14 March), 0.55--0.62 (15 March) and 0.93--0.00 (16 March). 

Given the inclination of the system (between $\sim70\degr$, 
Orosz~\&~Bailyn~\cite{oro97}, van~der~Hooft~et~al.~\cite{van98}; 
and $\sim85\degr$, Hjellming~\&~Rupen~\cite{hje95}), 
an orbital phase close to zero corresponds 
to having the donor star situated closer to the observer, while in phase 0.5 the disk 
would be in front of the secondary star. 
This means that if any of the observed spectral features were arising from the illuminated face 
of the secondary star, it would have its maximum at phase 0.5 and would not be present at phase zero.
This is not the case in our observations. 
Another possible orbital effect would be that the disk emission were absorbed by the 
stellar wind of the secondary star, which would produce increased absorptions 
close to phase zero, i.e. on 16 March, which is also not the case.

Finally, neither dips nor eclipses were observed in the disk thermal emission at any of the phases. 
Given the size of the donor star, $\sim5\rm{R}_{\odot}$ (Orosz~\&~Bailyn~\cite{oro97}), 
the binary separation ($1.17\times10^{12}$cm), and assuming that the soft X-ray emission
originates in the central 200\,000 km of the disk (see below, section~\ref{sect_fexxv}), 
the duration of a possible eclipse of the soft X-ray emission 
by the donor star would last more than 2.5 hours (i.e., a change 
in the orbital phase of 0.04) close to phase zero.
This should have been clearly visible during our 16 March observation. 

The fact that no orbital modulation is observed in the X-ray spectra, provides 
an upper limit for the inclination of the system.
With the parameters mentioned above for the sizes of the system, the inclination must be smaller than
73$\degr$ (for the innermost 200\,000~km of the disk surface to be visible in all orbital phases). 
This limit is in agreement with the inclination determined from the
optical light-curve by Orosz~\&~Bailyn~(\cite{oro97}), 
$69.\!\!^\circ5\pm0.\!\!^\circ8$, 
and by van~der~Hooft~et~al.~(\cite{van98}), $67.\!\!^\circ2\pm3.\!\!^\circ5$,
but incompatible with the inclination inferred from the radio jets, 
$\sim84\degr$ (Hjellming~\&~Rupen~\cite{hje95}).
This points, as suggested by Orosz~\&~Bailyn~(\cite{oro97}), to an inclination of the jet axis of 
about 15$\degr$ with respect to the normal to the orbital plane. 
Alternatively, non-symmetric jet ejections could have lead to a wrong inclination inference 
(Hjellming~\&~Rupen~\cite{hje95}).

\subsection{\label{sect_fexxv}The Fe~XXV/Fe~XXVI absorber}

Clear highly ionized Fe absorption lines are detected in the EPIC-pn spectra of 
the three March observations in the high/soft state.
It is worth noting that Ueda~et~al.~(\cite{ued98}) found simultaneously Fe~XXV absorption features 
and an iron edge, while they detected no edge when absorption in the iron K band corresponded to Fe~XXVI. 

Assuming that the absorbing plasma is photoionized, the ionization state of the observed 
elements provides information on the conditions and location of the absorber.
The presence of He-like Fe ions indicates an ionization parameter 
$\xi=$L/nr$^{2}\sim10^{3}$~erg~cm~s$^{-1}$ (Kallman~et~al.~\cite{kal96}).
The lack of lower ionization absorption implies that the absorber cannot have a large extent.

The equivalent width of the Fe lines is increasing 
between the first and the last March observations. 
This may be pointing out either an increase in the total column density,
or in the ionization state of the gas.

Ueda~et~al.~(\cite{ued98}) determined the iron column density of the plasma 
to be $10^{19}-10^{20}$~cm$^{-2}$ from the observed equivalent width
and the curve of growth of the Fe~XXV~K$\alpha$ line, 
which relates the expected equivalent width to the iron column density. 
Using their curve of growth and assuming the detected features are only Fe~XXV,
our equivalent width of the Fe~XXV~K$\alpha$ line (between 50 and 160~eV)  
corresponds to a column density of $10^{19}$ and $5\times10^{20}$~cm$^{-2}$, 
which is similar to the values found by Ueda~et~al.~(\cite{ued98}).
Assuming cosmic abundances, this corresponds to a hydrogen column density 
in the range $2\times10^{23}-10^{25}$ cm$^{-2}$. Since the observed absorption 
lines may be a blend of Fe~XXV and Fe~XXVI, these must be taken as upper limits.
In addition, for a hydrogen column density 
larger than 10$^{24}$ cm$^{-2}$, the line absorber would be optically thick to 
Thomson scattering (Ueda~et~al. 1998). We consider this to be unlikely and, 
taking into account the errors in the equivalent widths, 
the column density could be then in the range $2\times10^{23}- 10^{24}$~cm$^{-2}$
in all three observations.

With the photoionization parameter, $\xi\sim10^{3}$~cm~s$^{-1}$, 
and the flux detected above $\sim9$~keV (X-rays photoionizing He-like iron ions), 
$\rm L_{\ge 9\rm{keV}}\sim5\times10^{36}$~erg~s$^{-1}$ (for a distance of 3.2~kpc), 
the column density indicates a distance to the central source 
between 50\,000 and 200\,000~km. 
Assuming that the disk radius 
is 70\% of the Roche lobe radius, i.e., $4\times10^{6}$~km, 
the Fe~XXV absorber extends to less than 5\% of the disk surface.

In the case that the features correspond to blueshifted Fe~XXV lines, the 
error in the energy determination is too large as to reach any strong conclusion 
about its blueshift. 
Nevertheless, assuming the wind velocity is constant,  
the common range of blueshifts of the three March observations  
is reduced to 2600--4500~km~s$^{-1}$.
From the parameters derived above for the Fe~XXV absorber, 
and assuming constant density and spherical symmetry for the expanding wind, 
the mass loss rate $4\pi \rm r^2 \rho\rm v \rm b$ would be in the range 
$(2-13)\times10^{-7}\rm b ~\rm M_{\odot} \rm {yr}^{-1}$, with b being the filling factor. 
This upper limit is a factor 20--130 larger than the accretion rate, which would indicate a highly non-stationary situation.
However, there is considerable uncertainty in the wind mass loss, since both the wind density and wind
velocity are likely a function of radius and the wind may be conical rather than spherically symmetric. 
Exploring these effects goes beyond the scope of this paper.

\section{Summary}

GRO~J1655-40 was observed with XMM-Newton during the low/hard state in February 2005 and in the 
soft/high state in March 2005. The continuum spectrum is dominated by an absorbed power law in the 
low/hard state and by a multicolor disk blackbody in the soft/high state, with the 
maximum temperature increasing during the high state.
In addition, the X-ray spectrum during the March observations shows several spectral features:

- Interstellar neutral oxygen causes an OI~K$\alpha$ absorption line at 23.5~$\rm \AA$. 
A correlation between the equivalent width of this line and the hydrogen column density 
has been obtained from published observations of 
several galactic X-ray sources, and the comparison of our observations with 
that correlation confirms the interstellar origin of the OI~K$\alpha$ absorption line. 

- Fe~XXV, possibly blended with Fe~XXVI, 
is detected in absorption in the EPIC-pn spectra, indicating the presence of 
a highly ionized absorber, with an ionization parameter 1000~erg~cm~s$^{-1}$, 
extending less than 200\,000~km from 
the central source, corresponding to a 5\% of the accretion disk.

Photoionized absorbers are seen in AGN as well, also with expansion velocities. The detection of
photoionized winds in microquasars is then another similarity between these two kinds of systems.
Nevertheless, winds in AGN are not uniform, but rather distributed in clouds of material. If this
were the case in GRO~J1655-40, the absorption systems would go in and out of the line of sight and
cause temporal variations in the absorption features, which is not observed in our XMM-Newton observations.

\begin{acknowledgements}
The XMM-Newton project is an ESA Science Mission with instruments
and contributions directly funded by ESA Member States and the
USA (NASA). The XMM-Newton project is supported by the
Bundesministerium f\"ur Wirtschaft und Technologie/Deutsches Zentrum
f\"ur Luft- und Raumfahrt (BMWI/DLR, FKZ 50 OX 0001), the Max-Planck
Society and the Heidenhain-Stiftung.
We acknowledge the RXTE/ASM team for making quick-look results available for public use. 
We thank Norbert Schulz for pointing us the interstellar origin of the Fe L edges.
We acknowledge Marcus Kirsch, Michael Freyberg, Konrad Dennerl and Cor de Vries for
their help with EPIC-pn and RGS calibration issues. We also thank Y. Ueda for 
useful discussions. GS is supported through a postdoctoral fellowship 
from the Spanish Ministery for Education and Science.
XLZ is supported through DLR/FKZ 50 OX 0502.
\end{acknowledgements}

\section*{Note added in proof:}

Within a few days of placing the refereed accepted version 
of this paper on astro-ph, two other papers appeared about the same source 
GRO~J1655-40, which partially affect our conclusions and are commented on below:

\begin{enumerate}
\item{Foellmi~et~al.~(\cite{foe06}) have recently obtained a new determination for the distance to 
GRO~J1655-40 and find an upper limit of 1.7~kpc. In this case, the luminosity in the 
range 0.4--10~keV listed in Table \ref{tab_mod} would be $1.0\times10^{35},  9.6\times10^{36},  9.6\times10^{36}$ and
 $1.2\times10^{37}$~erg\,s$^{-1}$ (for days 27 Feb, 14, 15 and 16 March respectively), 
and the disk inner radius would be only 7.6~km, 7.0~km and 7.2~km (for days 14, 15 and 16 March 2005).
We note, however, that these values are smaller than the gravitational radius of the black hole, 
$\rm R_{\rm g}=\rm{GM/c}^2\sim10$~km. Therefore, either the distance to GRO~J1655-40 is 
at least 2.5~kpc, or the model for deriving the inner radius of the accretion 
disk requires improved modifications.}

\item{Miller~et~al.~(\cite{mil06}) reported on a Chandra observation of GRO~J1655-40 
performed on 1 April 2005, two weeks after our last XMM-Newton observation. 
The HETGS spectrum shows 90 absorption lines significant at the 5$\sigma$ level 
of confidence or higher. From all of these, only the Fe~XXV K$\alpha$ and K$\beta$ lines are also
present in our March observations. 
We have computed the upper limit for the flux of some of the strongest lines detected by 
Chandra in the RGS energy range, in particular, for the Fe XXIV 5p, 4p and 3p lines at 
7.159, 7.979, 10.604, 10.649 $\rm \AA$.
Fixing the energy and FWHM of the lines to the values 
obtained by Miller~et~al.~(\cite{mil06}), we obtain the following upper limits at 3$\sigma$ 
confidence level 
for the flux of the lines: 4.2, 0.36, 1.4 and 1.3($\times10^{-3}\rm{ph}\,\rm{cm}^{-2}\rm{s}^{-1}$), 
for the observation on 15 March 2005. 
These upper limits are between a factor 3 and 30 smaller than the fluxes measured in the 
Chandra spectrum and suggests a change in the ionization state between the two observation periods.}

\end{enumerate}

\appendix
\section{EPIC-pn CTE correction}

\begin{figure}
\resizebox{\hsize}{!}{\includegraphics[angle=-90]{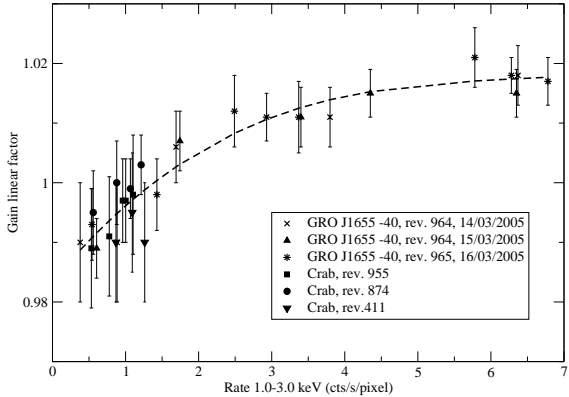}}
\caption{
Calibration of the gain linear factor \textit{f} as a function 
of rate per pixel \textit{r}.
\label{fig_cte}}
\end{figure}

An inaccurate calibration of the EPIC-pn charge transfer efficiency (CTE) 
leads to a slightly wrong energy determination of a detected photon, 
which becomes evident in the large residuals around sharp features like the instrumental Si and Au edges.
From our GRO~J1655-40 observations, the shift in energy has been found to be rate dependent, 
being stronger at the center of the PSF, 
which implies that it can not be directly corrected in the extracted spectrum.
We determined the improvement of the CTE (formally handled as a linear gain factor) 
for different rates, selecting and evaluating the energy gain linear factor for 
spectra extracted from different regions of the detector  (Fig.\ref{fig_cte}). 
Crab calibration observations in the Burst mode were also included to improve 
the determination of the dependence of the gain linear factor \textit{f} with 
the rate per pixel \textit{r} (cts/s/pixel), which we find can be approximated by 
$\rm f=0.98+0.015r-2.2\times10^{-3}\rm r^{2}+1.1\times10^{-4}\rm r^{3}$.

We used this linear gain to correct the energy of our event files:
for each photon, the PI channel was divided by the gain factor \textit{f} 
according to its rate \textit{r}, assigned to each photon depending on the RAWX column.
To avoid the discretization resulting from the pixelization of the PSF in the RAWX columns, 
the \textit{r} values were randomized within each RAWX column.

After correcting the event tables with this linear gain, no more 
residuals appear around the Au and Si edges in the new extracted spectra.
We confirmed that the correction is also valid at the high energy range, by checking
that, after the correction, the spectral features at 7--8~keV have the same central energy 
both at the center of the PSF (at high rate) and at the wings (low rate).

\end{document}